\documentclass[prd,twocolumn,showpacs,amsmath,amssymb,superscriptaddress,preprintnumbers,nofootinbib,showkeys]{revtex4}

\begin{document}

\newcommand{\nablab}{{\mathop {\rule{0pt}{0pt}{\nabla}}\limits^{\bot}}\rule{0pt}{0pt}}

\title{Axionic extension of the Einstein-aether theory}

\author{Alexander B. Balakin}
\email{Alexander.Balakin@kpfu.ru} \affiliation{Department of
General Relativity and Gravitation, Institute of Physics, Kazan
Federal University, Kremlevskaya str. 18, Kazan 420008,
Russia}

\date{\today}

\begin{abstract}
We extend the Einstein-aether theory to take into account the interaction between a pseudoscalar field, which describes the axionic dark matter, and a time-like dynamic unit vector field, which characterizes the velocity of the aether motion. The Lagrangian of the Einstein-aether-axion theory includes cross-terms based on the axion field and its gradient four-vector, on the covariant derivative of the aether velocity four-vector, and on the Riemann tensor and its convolutions.  We follow the principles of the Effective Field theory, and include into the Lagrangian of interactions all possible terms up to the second order in the covariant derivative. Interpretation of new couplings is given in terms of irreducible parts of the covariant derivative of the aether velocity,
namely, the acceleration four-vector, the shear and vorticity tensors,
and the expansion scalar. A spatially isotropic and homogeneous cosmological model with dynamic unit vector field and axionic dark matter is considered as an application of the established theory;
new exact solutions are discussed, which describe models with a Big Rip, Pseudo Rip and de Sitter-type asymptotic behavior.

\end{abstract}
\maketitle

\section{Introduction}

The Einstein-aether-axion theory can be classified as a pseudoscalar-vector-tensor theory of gravitation.
There are two intrinsic elements in this theory, which allow us to consider it as an alternative theory of gravity.
The first element is a dynamic time-like unit vector field $U^i$ characterizing the velocity of the aether (see, e.g., \cite{J1,J2,J3,J4,J5,J6,J7});
this unit vector  field realizes the idea of a preferred frame of reference (see, e.g., \cite{CW,N1,N2,N3}).
The second element is the pseudoscalar field associated with axionic dark matter (see, e.g., \cite{PQ,Weinberg,Wilczek,ADM1,ADM2,ADM3,ADM4,ADM5,ADM6,ADM7,ADM8} for review, references and mathematical details).

Formally speaking, the structure of this theory is similar to the one of  a scalar-vector-tensor theory of gravity, however, there is a few features, which distinguish them.
Three typical ways to extend the vector-tensor theory of gravity in order to obtain a scalar-vector-tensor version of this theory are known. The first way is standard, it is based
on introduction of three independent fields: scalar field $\Psi$, vector field $V^i$ and tensor field $g_{ik}$ associated with metric of the corresponding space-time (see, e.g., \cite{TeVeS,CW2} for details of the so-called TeVeS theories, and \cite{EAS1,EAS2} for discussion of the cosmological vector field coupled to a scalar inflaton). The second way is associated with introduction of a time-like four-vector field $B^i$, the scalar square of which, $B^kB_k \equiv {\cal B}^2$ gives the scalar function, so that the unit four-vector $U^i \equiv \frac{B^i}{{\cal B}}$ relates to the velocity of an aether. In other words, four degrees of freedom associated with the four-vector $B^i$ are redistributed between one scalar and three vectorial degrees of freedom. One can find fingerprints of this approach, for instance, in the so-called bumblebee models (see, e.g., \cite{Bumb1,Bumb2}).  The third way is admissible, when the unit time-like four-vector field $U^i$ has a quasi-gradient structure, i.e., it can be presented as $U_i = \frac{1}{X} \nabla_i \Phi$, where $X$ satisfies the relation $X^2= \nabla_m \Phi \nabla^m \Phi$, thus providing the normalization condition $U^mU_m=1$. In this particular case a new intrinsic scalar field $\Phi$, the so-called Khronon \cite{Khronon1}, appears in the theory, thus introducing a system of global causal time-like surfaces $\Phi{=}const$. This finding allowed the authors of \cite{Khronon1,Khronon2} to open a discussion concerning the so-called universal horizon, which acts as an one-way membrane, so that particles with arbitrary speed are trapped inside some spherical domain and cannot escape from it. This theoretical prediction has an interesting interpretation in terms of theory known as Analogue Black Holes (see, e.g., \cite{Analog1,Analog2} and references therein), and gives a possibility to speak about a new interpretation of dark energy phenomenon (see, e.g., \cite{Khronon3}).

The Einstein-Maxwell-aether theory \cite{BL2014} was elaborated as an extension, based on introduction of the Maxwell field into the Lagrangian.
The Einstein-Maxwell-aether theory operates, in fact, with two vector fields (to be more precise, one vector field and one co-vector field): the first one is the unit dynamic vector field $U^i$; the second is presented by the electromagnetic potential four-(co)vector, $A_i$. Modus operandi of the field $U^i$  differs essentially from the one of $A_i$. Indeed, the unit four-vector $U^i$ itself, as well as, both symmetric and skew-symmetric constituents of the covariant derivative $\nabla_k U_i$ enter the decomposition of the Lagrangian. On the contrary, only the skew-symmetric part of the derivative $F_{ik}=\nabla_iA_k{-}\nabla_kA_i$, the Maxwell tensor, enters the Lagrangian of the  Einstein-Maxwell-aether theory \cite{BL2014}; the four-vector $A_i$ itself does not appear in the Lagrangian thus providing the $U(1)$-gauge invariance of the theory.
The study of coupling of electromagnetic fields to the axionic dark matter activated the interest to the axion electrodynamics, which is the axionic extension of the Faraday-Maxwell theory (see, e.g., \cite{Ni77,Sikivie83,Wilczek82,Hehl,BG2013,BN2014}). In the framework of this axion electrodynamics, interesting results are obtained, which describe anomalous electromagnetic response of an axionically active cosmic medium dynamo-optically coupled to the unit vector field $U^i$ (see, e.g., \cite{AB2016}).

In this context, the establishing of the full-format Einstein-aether-axion model can be considered as a twofold task. First, when one deals with cosmological applications of the Einstein-aether theory, it seems to be reasonable to consider the interaction of the dark matter with a non-uniformly moving aether; if the dark matter contains an axionic fraction, this coupling is the subject of the Einstein-aether-axion theory. Second, the Einstein-aether-axion model is a missing element, which is necessary to complete the Einstein-Maxwell-aether-axion theory. Indeed, photons interact directly with the pseudoscalar field (direct axion-photon coupling), and with the unit vector field (direct dynamo-optical coupling). In addition, one can search for indirect photon couplings, mediated by an axion-aether interactions, and this mediation is the subject of the Einstein-aether-axion theory.

One remark  should be made concerning the links between scalar and pseudoscalar extensions of the Einstein-aether theory. The invariant terms in the Lagrangian, which contain the pseudoscalar
field $\phi$ in an even degree (e.g., $\phi^2$, $\phi \nabla_i \phi$, $\nabla_i \phi \nabla_k \phi$) do not differ, formally speaking, from the corresponding terms containing the true scalar field $\Phi$. However, when the invariant cross-terms in the Lagrangian of these models contain pseudoscalar and scalar fields in the linear form, the essential difference between pseudoscalar and scalar  extensions of the Einstein-aether theory readily appears. Clearly, in case of pseudoscalar field such linear terms have to include the products $\phi \epsilon^{ikmn}$ and $\nabla_s \phi \epsilon^{ikmn}$ with (pseudo)tensor Levi-Civita $\epsilon^{ikmn}$, providing all the terms in the Lagrangian to be true scalar invariants. In other words, the complete axionic extension of the Einstein-aether theory is not equivalent to the scalar extension of this theory.

In this paper we establish the version of the Einstein-aether-axion theory, which can be classified as the theory of the second order with respect to the Effective Field theory.
(see, e.g.,  \cite{EFT1,EFT2,EFT3,EFT4,EFT5} for details, terminology and historical remarks).
Mathematically, we extend the Lagrangian by inserting all possible cross-terms,
which, on the one hand, contain the vector field $U^i$, and the tensor $\nabla_i U_k$, the covariant derivative of the unit vector field,
and, on the other hand, include pseudoscalar (axion) field $\phi$ and its gradient four-vector $\nabla_i \phi$. The reconstructed Lagrangian is up to the second order in the covariant derivative $\nabla_k$. This means that $\phi$ and $U^i$ can enter arbitrarily; the Riemann tensor $R^i_{\ kmn}$, the Ricci tensor $R_{ik}$ and Ricci scalar $R$ enter linearly; the derivatives of $\phi$ and $U_i$ appear only in the following forms: ($\nabla_i \phi$), ($\nabla_i U_k$), ($\nabla_i \phi \cdot \nabla_k \phi$),
($\nabla_i \phi \cdot \nabla_k U_j$), ($\nabla_i U_k \cdot \nabla_l U_s$). Other constructive elements used in the Lagrangian are the metric, Kronecker deltas and Levi-Civita tensor. Coupling constants are introduced phenomenologically; in order to classify them adequately we use the decomposition of of the tensor $\nabla_iU_k$  with
respect to its irreducible parts, namely, the acceleration four-vector, the shear and vorticity tensors, and the expansion
scalar.

The paper is organized as follows. In Section II, we reconstruct the Lagrangian classifying and listing all the appropriate invariant terms. In Section III we derive master equations for pseudoscalar, unit vector and gravitational fields. In Section IV we consider cosmological applications  of the established model. Section V includes discussion and conclusions.

\section{The Einstein-aether-axion theory: Reconstruction of the action functional}

The Einstein-aether-axion theory can be described by the total action functional
$$
S_{({\rm EAA})} = \int d^4 x \sqrt{{-}g} \pounds \,,
$$
\begin{equation}
\pounds \equiv L_{({\rm G})} + L_{({\rm U})} + L_{({\rm A})} + L_{({\rm cross})} \,,
\label{0}
\end{equation}
where $g$ is, as usual, the determinant of the metric tensor $g {=} {\rm det}(g_{ik})$. The total Lagrangian $\pounds$ includes the Lagrangian of the gravity field $L_{({\rm G})}$; the Lagrangian of the unit dynamic vector field (the aether velocity four-vector) $ L_{({\rm U})}$; the Lagrangian of the pseudoscalar (axion) field $L_{({\rm A})}$, and the Lagrangian of interactions $L_{({\rm cross})}$. Let us discuss them in more details.

\subsection{Lagrangian of the pure Einstein-aether theory}

For the Einstein-Hilbert action, the Lagrangian is well-known, $L_{({\rm G})}= \frac{1}{2\kappa}\left[R{+}2\Lambda \right]$; here $R$ is the Ricci
scalar, $\Lambda$ is the cosmological constant, and $\kappa {=} 8\pi G$, with the Newtonian coupling constant $G$ ($c{=}1$ in the chosen units). The term $L_{({\rm U})}$ describes the Lagrangian of the dynamic vector field $U^i$ (see, e.g., \cite{J1}):
\begin{equation}
L_{({\rm U)}} {=} \frac{1}{2\kappa}  \left[\lambda \left(g_{mn}U^m
U^n {-}1 \right) {+} K^{abmn} \nabla_a U_m  \nabla_b U_n
\right].
\label{1}
\end{equation}
The first term $\lambda \left(g_{mn}U^m U^n {-}1 \right) $ ensures
that the $U^i$ is normalized to one, and $\lambda$ is the Lagrange multiplier.
The second term $K^{abmn} \ \nabla_a U_m \ \nabla_b U_n $ is quadratic in the covariant derivative
$\nabla_a U_m $ of the vector field $U^i$, with the tensor $K^{abmn}$ to be constructed
using the metric tensor $g^{ij}$ and the aether velocity four-vector $U^k$ only,
\begin{equation}
K^{abmn} {=} C_1 g^{ab} g^{mn} {+} C_2 g^{am}g^{bn}
{+} C_3 g^{an}g^{bm} {+} C_4 U^{a} U^{b}g^{mn}.
\label{2}
\end{equation}
The parameters $C_1$, $C_2$, $C_3$ and $C_4$ are the Jacobson's coupling constants \cite{J1}.

It is well-known that the tensor $\nabla_i U_k$ can be decomposed into a sum of
its irreducible parts, namely, the acceleration four-vector $DU^{i}$,
the shear tensor $\sigma_{ik}$, the vorticity tensor $\omega_{ik}$, and
the expansion scalar $\Theta$, as follows:
\begin{equation}
\nabla_i U_k = U_i DU_k + \sigma_{ik} + \omega_{ik} +
\frac{1}{3} \Delta_{ik} \Theta \,. \label{act3}
\end{equation}
The basic quantities in this decomposition are defined as
$$
DU_k \equiv  U^m \nabla_m U_k \,, \quad \sigma_{ik}
\equiv \frac{1}{2}\left(\nablab_i U_k {+}
\nablab_k U_i \right) {-} \frac{1}{3}\Delta_{ik} \Theta  \,,
$$
$$
\omega_{ik} \equiv \frac{1}{2} \left(\nablab_i U_k {-} \nablab_k U_i \right) \,, \quad \Theta \equiv \nabla_m U^m
\,,
$$
\begin{equation}
D \equiv U^i \nabla_i \,, \quad \Delta^i_k = \delta^i_k - U^iU_k \,, \quad \nablab_i \equiv \Delta_i^k \nabla_k \,. \label{act4}
\end{equation}
In these terms the scalar $K^{abmn}(\nabla_a U_m) (\nabla_b U_n)$ has the form
$$
K^{abmn}(\nabla_a U_m) (\nabla_b U_n) =
$$
$$
=(C_1 {+} C_4)DU_k DU^k {+}
(C_1 {+} C_3)\sigma_{ik} \sigma^{ik} {+}
$$
\begin{equation}
+ (C_1 {-} C_3)\omega_{ik}
\omega^{ik} {+} \frac13 \left(C_1 {+} 3C_2 {+}C_3 \right) \Theta^2
\,. \label{act5}
\end{equation}

\subsection{Lagrangian of a pseudoscalar (axion) field}

The term $L_{({\rm A})}$ relates to the axionic dark matter:
\begin{equation}
L_{({\rm A})} = \frac{1}{2}\Psi^2_0  \left[V(\phi^2)  {-}  \xi R \phi^2 {-} g^{mn} \nabla_m \phi \nabla_n \phi
\right] \,.
\label{3}
\end{equation}
Here  the dimensionless quantity $\phi$ denotes the pseudoscalar (axion) field. The parameter $\Psi_0$ is reciprocal
to the constant of the axion-photon coupling $g_{({\rm A} \gamma \gamma)} {=}\frac{1}{\Psi_0}$. The coupling parameter $g_{({\rm A} \gamma \gamma)}$ can be presented as
$g_{({\rm A} \gamma \gamma)}{=}\frac{\alpha \zeta}{\pi f_{\rm a}}$, where $f_{\rm a}$ is the energy scale of the symmetry breaking (axion decay constant); $\alpha$ is the fine structure constant and $\zeta$ is a dimensionless model-dependent factor (its value is about one). Recently, the CAST Collaboration (CERN Axion Solar Telescope) reported that the improved limit for the  constant of the axion-photon coupling is $g_{({\rm A} \gamma \gamma)}< 1.47 \times 10^{-10} {\rm GeV}^{-1}$ (see \cite{CAST14}).
The parameter $\xi$ is the constant of nonminimal coupling of axion and gravitational fields; $V(\phi^2)$ is the potential of the pseudoscalar field.
We do not include "derivative coupling" terms containing the Ricci scalar and Ricci tensor in combinations with the gradient
four-vectors $\nabla_k \phi$ (say, $R^{ik} \nabla_i \phi \nabla_k \phi$, see, e.g., \cite{NM5}), since according to the Effective Field theory approach, such terms would be of the fourth order in the covariant derivative.
As usual, we omit the terms proportional to quantities  $\nabla^m \nabla_m \phi^2$, $\nabla^m [\phi  \nabla_m \phi]$, $\phi \nabla^m \nabla_m \phi$, which can be transformed into terms already written or into the perfect four-divergence, when we deal with the integration by parts in the action functional.
One can remark that the aether velocity $U^i$ does not enter this part of the Lagrangian directly; however,
using the decomposition
\begin{equation}
\nabla_i \phi \equiv U_i D\phi + \nablab_i \phi
\label{57}
\end{equation}
one can divide (when it is necessary) the gradient four-vector $\nabla_i \phi$ into the part proportional to the convective derivative $D\phi=U^s\nabla_s \phi$, and the part containing the spatial gradient   $\nablab_i \phi {=} \Delta_i^k \nabla_k \phi$ defined in terms of preferred frame of reference, associated with the aether velocity.

\subsection{Classification of interaction terms}

Following the principles of Effective Field theory \cite{EFT1,EFT2,EFT3,EFT4} one can display all the invariant terms up to second orders
in the derivatives in the context of the Einstein-aether-axion model.
We intend to show that there are two zero-order terms (we mark them as (0.1) and (0.2), respectively);
one irreducible type of the first order terms (1); four irreducible types
of the second order terms (2.1), (2.2), (2.3) and (2.4), respectively.
Let us list possible cross-invariants and distinguish irreducible terms.

\subsubsection{Terms of zero order in derivatives}

\vspace{3mm}
\noindent
{\it The term indicated as (0.1)}

\noindent
Typical terms of this type contain the squared pseudoscalar (axion) field and have the form $f(\phi^2)$, where $f$ is arbitrary function of its argument.
One of such terms is included into the Lagrangian of the pseudoscalar field (\ref{3}) as the corresponding potential $V(\phi^2)$.

\vspace{3mm}
\noindent
{\it The term indicated as (0.2)}

\noindent
There is one term of this type involving the
aether velocity $U^i$. It is proportional to $g_{ik}U^i U^k$, and is already included
into the Lagrangian of the unit dynamic vector field $L_{({\rm U})}$.

\subsubsection{Terms of the first order in derivatives}

The terms of the first order in derivatives can be divided into three subtypes, nevertheless, only one of them can be included into the set of irreducible invariant terms.

\vspace{3mm}
\noindent
{\it The terms of the type (1.1)}.

\noindent
This first subtype includes the covariant derivative of the vector field and does not contain $\phi$; the terms of this type have the form
$\alpha^{ik} \nabla_i U_k$.
We assume that the tensorial coefficients $\alpha^{ik}$ can be constructed using the
metric $g_{ik}$, the Kronecker tensors
($\delta^i_k$, $\delta^{ik}_{ab}$ and higher order Kronecker tensors),
the Levi-Civita tensor $\epsilon^{ikab}$, and the unit vector field
$U^k$. The metric $g_{ik}$, the Kronecker deltas, the Levi-Civita (pseudo) tensor are covariantly constant, i.e., $\nabla_s g_{mn}{=}0$, $\nabla_s \delta^{ab}_{mn}{=}0$, $\nabla_s \delta^{a}_{b}{=}0$, $\nabla_s \epsilon^{abmn}{=}0$.
Clearly, this subtype contains only one appropriate scalar, namely, $\alpha \Theta$, where $\Theta{=}\nabla_kU^k$ is
the expansion scalar, and $\alpha$ is a coupling constant
introduced phenomenologically. The corresponding term in the action functional, $\sqrt{-g}\alpha \Theta$,  can be reduced to the perfect divergence  $\sqrt{-g}\alpha \Theta {=}  \partial_k (\alpha \sqrt{-g}U^k)$, and we omit the term $\alpha^{ik} \nabla_i U_k$ from our consideration.

\vspace{3mm}
\noindent
{\it The terms of the type (1.2)}.

\noindent
The cross-terms of the form $(\nabla_kU^k) f(\phi^2)$ are extensions of terms mentioned in the item (1.1); now it is irreducible and can not be discarder.

\vspace{3mm}
\noindent
{\it The terms of the type (1.3)}.

\noindent
There is also the invariant cross-term $ \phi g(\phi^2)U^k \nabla_k \phi$, which  contains the convective derivative of the pseudoscalar field. Due to the relationship
\begin{equation}
(\nabla_kU^k) f(\phi^2) = \nabla_k [U^k f(\phi^2)] - 2 \phi f^{\prime}(\phi^2)U^k \nabla_k \phi
\end{equation}
this contribution can be reduced to (1.2), and to a perfect divergence, using the integration by parts in the action functional.

\subsubsection{Terms of the second order in derivatives}

There are four subtypes in the list of terms of the second order in derivatives.

\vspace{3mm}
\noindent
{\it The terms of the type (2.1)}.

\noindent
The terms, which include covariant derivatives of the unit vector field and do not contain the pseudoscalar field $\phi$,  can be reduced to  the invariant
\begin{equation}
K^{abmn} \nabla_a U_m \nabla_b U_n\,,
\label{b1}
\end{equation}
which enters the Lagrangian $L_{({\rm U})}$.
It is well-known that other terms can be absorbed into this invariant.
Indeed, terms with second-order covariant derivatives $ {\cal
K}^{ikl}\nabla_i \nabla_k U_l$
can be rewritten as follows
\begin{equation}
{\cal K}^{ikl}\nabla_i \nabla_k U_l =\nabla_i \left[{\cal K}^{ikl}
\nabla_k U_l\right] - (\nabla_k U_l) \nabla_i \left({\cal
K}^{ikl}\right)\,.
\label{b2}
\end{equation}
Since the tensor $ {\cal K}^{ikl}$
contains the unit vector field
$U^k$, and the covariantly constant quantities: the metric $g_{ik}$, the Kronecker tensors ($\delta^i_k$,
$\delta^{ik}_{ab}$, etc.), the
Levi-Civita tensor $\epsilon^{ikab}$, we obtain that
\begin{equation}
{\cal K}^{ikl}\nabla_i \nabla_k U_l {=} \nabla_i \left[{\cal
K}^{ikl} \nabla_k U_l\right] {-} (\nabla_k U_l) (\nabla_i U^j)
\frac{\partial {\cal K}^{ikl}}{\partial U^j} \,.
\label{b3}
\end{equation}
The first term in the right-hand side of this relationship is a
perfect four-divergence, which can be omitted, and
the second term can be
included into $K^{abmn} \nabla_a U_m \nabla_b U_n$ by redefinition of
the tensor $K^{abmn}$.

Similarly, the nonminimal term $R_{ik}U^i U^k$ can be absorbed into the term $K^{abmn} \nabla_a U_m \nabla_b U_n$.
Indeed, using the relationship
\begin{equation}
[\nabla_a \nabla_b - \nabla_b \nabla_a ]U_m = - U^s R_{smab} \,,
 \label{b5}
\end{equation}
the term $R_{ik}U^i U^k$ can be rewritten as
$$
R_{ik}U^i U^k = \nabla_i \left[U^k \nabla_k U^i- U^i\nabla_k U^k
\right]+
$$
\begin{equation}
+\left(\nabla_i U^i \right)\left(\nabla_k U^k \right) -
\left(\nabla_m U^k \right)\left(\nabla_k U^m \right) \,.
 \label{b6}
\end{equation}
The first term is a perfect four-divergence,
and the other terms can be included in the construction of the
Jacobson's type term $K^{abmn} (\nabla_a U_m)(\nabla_b U_n)$. Here
and below we use the parentheses in the expressions of the form
$(\nabla_a U_m) {\cal T}$ just to indicate that the covariant
derivative operator acts on $U_m$ only.

\vspace{3mm}
\noindent
{\it The terms of the type (2.2)}.

\noindent
Terms, which include covariant derivatives of the unit vector field and arbitrary functions of $\phi^2$,  can be represented as
\begin{equation}
{\cal K}^{abmn} \nabla_a U_m \nabla_b U_n\,,
\label{b8}
\end{equation}
where
$$
{\cal K}^{abmn} = h_1(\phi^2) g^{ab} g^{mn} {+} h_2(\phi^2) g^{am}g^{bn}
{+} h_3(\phi^2) g^{an}g^{bm} {+}
$$
\begin{equation}
{+} h_4(\phi^2) U^{a} U^{b}g^{mn} {+} \phi h_5(\phi^2) \epsilon^{abmn} {+} \phi h_6(\phi^2) \epsilon^{asmn}U_s U^b\,.
\label{b9}
\end{equation}
The term ${\cal K}^{abmn}$ is reconstructed similarly to $K^{abmn}$ with functions $h_1(\phi^2)$, $h_2(\phi^2)$, $h_3(\phi^2)$, $h_4(\phi^2)$ instead of Jacobson's constants $C_1$, $C_2$, $C_3$ and $C_4$, respectively. Two last terms (containing the functions $h_5(\phi^2)$ and $h_6(\phi^2)$) appear due to presence of the pseudoscalar field $\phi$, since the new true tensor $\phi \epsilon^{abmn}$ can be now used as a constructive element of the Lagrangian. These are new irreducible interaction terms.

\vspace{3mm}
\noindent
{\it The terms of the type (2.3)}.

\noindent
The cross-terms, which include covariant derivative of the unit vector field and four-gradient of the pseudoscalar field $\phi$,  can be written as follows
\begin{equation}
{\cal H}^{abm} (\nabla_a U_m) (\nabla_b \phi) \,,
\label{b10}
\end{equation}
\begin{equation}
{\cal H}^{abm} =  \phi h_7(\phi^2) g^{am} U^b + \phi h_8(\phi^2) g^{bm} U^a
\,.
\label{b11}
\end{equation}
Clearly, the term proportional to $U^m \nabla_a U_m$ vanishes because of normalization of the unit vector field. Interaction terms given by (\ref{b10}), (\ref{b11}) are also irreducible .

\vspace{3mm}
\noindent
{\it The terms of the type (2.4)}.

\noindent
The cross-terms, which are quadratic in the gradient four-vector of the pseudoscalar field,  can be represented in the following form:
\begin{equation}
{\cal T}^{ab} (\nabla_a \phi) (\nabla_b \phi) \,,
\label{b12}
\end{equation}
where
\begin{equation}
{\cal T}^{ab} = h_9(\phi^2) g^{ab} + h_{10}(\phi^2) U^a U^b  \,.
\label{b13}
\end{equation}
The term with $h_9(\phi^2) \to -\frac{1}{2}\Psi^2_0$ was already used in the Lagrangian of the axion field. The term
\begin{equation}
h_{10}(\phi^2) U^a U^b (\nabla_a \phi) (\nabla_b \phi)
\end{equation}
is a principally new contribution, which appears in the presence of the unit vector field $U^i$.

There are few terms of the second order in derivatives, which can be reduced to the terms already listed above.

\vspace{3mm}
\noindent
{\it ${(a)}$
The term  $G^{abm}\nabla_b \nabla_a U_m$, which includes the second covariant derivative of the unit vector field, are not irreducible}.

\noindent
Keeping in mind the formula
$$
G^{abm}\nabla_b \nabla_a U_m  = - \left[\frac{\partial G^{abm}}{\partial U_n} \right] (\nabla_a U_m) (\nabla_b U_n)  -
$$
\begin{equation}
{-}\left[\frac{\partial G^{abm}}{\partial \phi}\right] (\nabla_a U_m) (\nabla_b \phi) {+} \nabla_b [G^{abm}\nabla_a U_m]\,,
\label{b14}
\end{equation}
one can see, that the last term is the perfect divergence; the first and second parts can be absorbed by the already listed irreducible terms of the type (2.2) and (2.3), respectively.

\vspace{3mm}
\noindent
{\it ${(b)}$
The extended nonminimal term}
$$
f(\phi^2) R_{ik}U^i U^k =
\nabla_i \{f(\phi^2) \left[U^k \nabla_k U^i- U^i\nabla_k U^k \right]\} -
$$
$$
-
 \left[U^k \nabla_k U^i- U^i\nabla_k U^k \right] \nabla_i f(\phi^2)+
$$
\begin{equation}
+ f(\phi^2)\left(\nabla_i U^i \right)\left(\nabla_k U^k \right) -
f(\phi^2) \left(\nabla_m U^k \right)\left(\nabla_k U^m \right)
\label{b15}
\end{equation}
can be, clearly, reduced to perfect divergence and to the terms of the form (2.2) and (2.3).

\vspace{3mm}
\noindent
{\it ${(c)}$
Invariants of the type ${\cal H}^{abm} (\nabla_a U_m) (\nabla_b \phi)$  with ${\cal H}^{abm} =f(\phi^2) \epsilon^{abmn}U_n$ are not irreducible.}

\noindent
Let us  make the transformations
$$
f(\phi^2) \epsilon^{abmn}U_n (\nabla_a U_m) (\nabla_b \phi) =
$$
$$
=\epsilon^{abmn}U_n (\nabla_a U_m) (\nabla_b F(\phi)) =
$$
$$
=\nabla_b \left[\epsilon^{abmn}U_n (\nabla_a U_m) F(\phi)\right] -
$$
\begin{equation}
- \epsilon^{abmn} F(\phi) \left[U_n (\nabla_b \nabla_a U_m) + (\nabla_b U_n) (\nabla_a U_m) \right] \,,
\label{b16}
\end{equation}
where $F(\phi)=\int d\phi f(\phi^2)$. The first term is, clearly, the perfect divergence; the last term can be absorbed by the invariants of the type (2.2). Keeping in mind the relationship (\ref{b5}), one can transform the term $\epsilon^{abmn} F(\phi) U_n (\nabla_b \nabla_a U_m)$ as follows:
$$
\epsilon^{abmn} F(\phi) U_n (\nabla_b \nabla_a U_m)=
$$
$$
=\frac12  F(\phi) U_n  \epsilon^{abmn} R^{s}_{\ mba} U_s =
$$
\begin{equation}
=\frac16  F(\phi) U_n  \epsilon^{abmn} R^{s}_{\ (mba)} U_s \,.
\label{b17}
\end{equation}
Since the cyclic transposition of indices in the Riemann tensor gives zero, $R^{s}_{\ (mba)} {=} 0$, this term vanishes. Thus, the cross-invariants of the type $f(\phi^2) \epsilon^{abmn}U_n (\nabla_a U_m) (\nabla_b \phi)$ are not irreducible.

\vspace{3mm}
\noindent
{\it ${(d)}$
The terms ${\cal G}^{ba}\nabla_b \nabla_a \phi$ are not irreducible.}

\noindent
Using the transformation
$$
{\cal G}^{ba}\nabla_b \nabla_a \phi  = \nabla_b [{\cal G}^{ba}\nabla_a \phi] -
$$
\begin{equation}
-\left[\frac{\partial {\cal G}^{ba}}{\partial \phi}\right] (\nabla_a \phi) (\nabla_b \phi)- \left[\frac{\partial {\cal G}^{ba}}{\partial U_n} \right] (\nabla_a \phi) (\nabla_b U_n) \,,
\label{b18}
\end{equation}
one can see that the first term in the right-hand side is the perfect divergence, and other terms can be reduced to the already listed terms of the types (2.4) and (2.3), respectively.

\subsubsection{A resume: Extension of the Lagrangian with the terms linear in the axion field $\phi$}

According to the classification of the interaction terms, one can extract from the Lagrangian $L_{({\rm cross})}$ the part $L_{(1)}$, which is linear in  the pseudoscalar (axion) field $\phi$. This part $L_{(1)}$ includes the skew-symmetric Levi-Civita (pseudo)tensor $\epsilon^{abmn}\equiv \frac{E^{abmn}}{\sqrt{-g}}$ (with $E^{0123}=1$) and is of the following form:
\begin{equation}
L_{(1)} =  \frac12 \Psi_0   \phi  \epsilon^{abmn}  [ h_5 (\nabla_a U_m)  {+} h_6 U_a D U_m](\nabla_b U_n) \,.
\label{4}
\end{equation}
Here $h_5$ and $h_6$ are phenomenological constants of linear interaction; they are zero-order realizations of the terms with $h_5(\phi^2)$ and $h_6(\phi^2)$ in (2.2).

\subsubsection{A resume: Extension of the Lagrangian with the terms quadratic in the axion field $\phi$}

When the pseudoscalar field $\phi$ enters in quadratic form, i.e., as
$\phi^2$, $\phi \nabla_k \phi$, $\phi \nabla_a \nabla_b \phi$, $\nabla_a ( \phi \nabla_b \phi)$, there are eight irreducible parts, which form the Lagrangian  $L_{({2})}$:
$$
 L_{({2})} = \frac12 \Psi^2_0  \left\{ \phi^2 \left[ \alpha \nabla_m U^m + h_1 (\nabla^a U^m)(\nabla_a U_m) + \right. \right.
 $$
$$
 \left. \left. + h_2 (\nabla_m U^m)^2 + h_3 (\nabla^a U_m)(\nabla^m U_a)
 +
  \right. \right.
$$
$$
\left. \left.
+ h_4 (U^a \nabla_a U_m)(U^b \nabla_b U^m)  \right]  + \right.
$$
$$
\left. + \phi \left[h_7 (\nabla_m U^m)(U^a \nabla_a \phi) + h_8 (U^a\nabla_a U^m)(\nabla_m \phi)\right]+ \right.
$$
\begin{equation}
\left. + h_{10} (U^a \nabla_a \phi)^2  \right\}
 \,.
\label{11}
\end{equation}
Now we are ready to write the total action functional for the Einstein-aether-axion theory in the second order version of the Effective Field theory and with up to the second order
terms in the pseudoscalar field.

\subsection{Total action functional}

The total action functional of the Einstein - aether - axion model of the second order with respect to the classification based on the Effective Field theory, and up to the second order with respect to axion field $\phi$ and its gradient four-vector $\nabla_k \phi$, can be written now in the form
$$
S_{({\rm EAA)}} =  \int d^4 x \sqrt{{-}g} \ \left\{ \frac{1}{2\kappa}\left[R{+}2\Lambda {+} \lambda (g_{mn}U^m
U^n {-}1 ){+} \right.\right.
$$
$$
\left. \left. +K^{abmn} (\nabla_a U_m) (\nabla_b U_n) \right] + \right.
$$
$$
\left. +\frac{1}{2}\Psi^2_0 \left[h_{10}(U^k\nabla_k \phi)^2 - g^{mn}\nabla_m \phi \nabla_n \phi \right]+ \right.
$$
$$
\left.{+} \frac{1}{2}\Psi^2_0 \phi^2 \left[m^2_{({\rm A})} {+} \alpha \nabla_k U^k {+} \left(h_1 g^{ab} g^{mn} {+} h_2 g^{am}g^{bn} {+} \right. \right. \right.
$$
$$
\left. \left. \left.
{+} h_3 g^{an}g^{bm} {+} h_4 U^{a} U^{b}g^{mn}\right) (\nabla_a U_m) (\nabla_b U_n) {-} \xi R \right] +
\right.
$$
$$
\left. {+} \frac12 \Psi_0   \phi  \epsilon^{abmn}  \left[h_5(\nabla_aU_m)(\nabla_bU_n) {+} \right.\right.
$$
$$
\left. \left.+ h_6 U_a  (U^k \nabla_k U_m) (\nabla_bU_n) \right] +
\right.
$$
$$
\left.
 + \frac12 \Psi_0^2 \left[h_7 (\nabla_m U^m ) (U^k \nabla_k \phi^2) + \right.\right.
$$
\begin{equation}
\left. \left.+  h_8 (U^k \nabla_k U^m)(\nabla_m \phi^2)\right]\right\}
\,.\label{total2}
\end{equation}
This representation of the total action functional of the Einstein-aether-axion model is not unique, however, any other representations can be reduced to this one using the integration by parts and redefinitions of the coupling constants introduced phenomenologically.

\section{Master equations for the pseudoscalar, vector, and gravitational fields}

\subsection{Master equation for the axion field}

Variation of the action functional $S_{({\rm EAA)}}$ (\ref{total2}) with respect to pseudoscalar field $\phi$ gives the linear equation
\begin{equation}
\nabla_m \left[G^{mn}\nabla_n \phi \right] + \phi {\cal H} = {\cal J} \,.
\label{ax1}
\end{equation}
Here the tensor
\begin{equation}
G^{mn} \equiv g^{mn}-h_{10} U^mU^n
\label{ax2}
\end{equation}
plays the role of acoustic metric for the pseudoscalar modes propagation, and thus the phenomenological parameter $h_{10}$
can be expressed through the velocity of pseudo-sound $V_{({\rm s})}$ as $h_{10}=1-\frac{1}{V^2_{({\rm s})}}$ (we use the units with $c{=}1$).
The term   ${\cal H}$ given by
$$
{\cal H} {=} m^2_{({\rm A})} {+} \alpha \Theta {+} (h_1{+} h_4) DU_m DU^m  {+} (h_1 {+} h_3 {-}h_8)\sigma_{ik} \sigma^{ik} {+}
$$
$$
+ (h_1 {-} h_3 {+}h_8)\omega_{ik}
\omega^{ik} + \frac13\left(h_1{+}3 h_2{+}h_3 {-}3h_7 {-}h_8 \right)\Theta^2  -
$$
\begin{equation}
- (h_7{+}h_8) D\Theta - \xi R -h_8  R_{mn} U^m U^n \,,
\label{ax3}
\end{equation}
plays the role of the squared effective mass, which is not constant and depends on the state of unit vector field.
The pseudoscalar ${\cal J}$
\begin{equation}
{\cal J} \equiv
\frac{1}{\Psi_0}\left[h_5 \omega^{am} +(2h_5{+}h_6) U^a  DU^m \right]\omega^{*}_{am}
\label{ax4}
\end{equation}
plays the role of the source of the axion field, which is produced by the unit vector field; clearly, this source is non-vanishing if and only if $\omega_{am} \neq 0$, i.e., the velocity field is characterized by non-vanishing vorticity tensor.

\subsection{Equations for the unit dynamic vector field}

The aether dynamic equations can be found by varying the action (\ref{total2}) with
respect to the Lagrange multiplier $\lambda$ and to the unit vector field $U^i$.
The variation of the action (\ref{total2}) with respect to
$\lambda$ yields the equation
\begin{equation}
g_{mn}U^m U^n = 1 \,,
\label{21}
\end{equation}
which is known to be the normalization condition of the time-like
vector field $U^k$.
Then, variation of the functional (\ref{total2}) with respect to
$U^i$ yields that $U^i$ itself satisfies the standard equation
\begin{equation}
\nabla_a {\cal J}^{aj}
- I^j = \lambda \ U^j  \,.
\label{0A1}
\end{equation}
Here the following nomenclature is used. First, the non-symmetric tensor ${\cal J}^{aj}$ is defined as the sum
\begin{equation}
{\cal J}^{aj} = {\cal J}^{aj}_{({\rm U})} +  \kappa \left[{\cal J}^{aj}_{(\rm h)}+{\cal J}^{aj}_{(\phi)} \right] \,,
\label{J1}
\end{equation}
displaying three intrinsic elements
\begin{equation}
{\cal J}^{aj}_{({\rm U})} = K^{abjn} (\nabla_b U_n)  \,,
\label{J2}
\end{equation}
\begin{equation}
{\cal J}^{aj}_{({\rm h})} =  \Psi_0^2 \phi^2 {\cal H}^{abjn} (\nabla_b U_n) \,,
\label{J2R}
\end{equation}
\begin{equation}
{\cal J}^{aj}_{(\phi)} {=}  \Psi_0 \phi \left[h_5 \epsilon^{abjn} {+}  h_6 U_s U^{[a} \epsilon^{b]jsn}  \right](\nabla_b U_n) \,,
\label{J4}
\end{equation}
where the tensor $K^{abjn}$ is given by (\ref{2}), and the new tensor ${\cal H}^{abjn}$ is introduced as
\begin{equation}
{\cal H}^{abjn} = \left[h_1 g^{ab} g^{jn} {+} h_2 g^{aj}g^{bn}
{+} h_3 g^{an}g^{bj} {+} h_4 U^{a} U^{b}g^{jn} \right] \,,
\label{J3}
\end{equation}
in analogy with (\ref{2}). Second, we introduced the vectors
\begin{equation}
I^j = I^j_{({\rm U})} + \kappa I^j_{(\phi)}\,,
\label{J5}
\end{equation}
\begin{equation}
I^j_{({\rm U})} =  C_4 (DU_m)(\nabla^j U^m) \,,
\label{J6}
\end{equation}
$$
I^j_{(\phi)} =  \Psi^2_0 \phi^2 h_4(DU_m)(\nabla^j U^m) + \Psi^2_0 \nabla^j \phi \left(h_{10} D \phi {-}\alpha \phi \right) +
$$
$$
+\Psi_0 \phi h_6 \left[DU_m \omega^{*mj} + (\nabla^j U_m) \omega^{*ma}U_a\right] +
$$
$$
+ \frac12\kappa \Psi^2_0 (h_7{+}h_8)D (\nabla^j \phi^2) +
$$
\begin{equation}
 + \frac12\kappa \Psi^2_0 (h_7{-}h_8)\left[\Theta (\nabla^j \phi^2) - (\nabla^j U^m)(\nabla_m \phi^2)\right]  \,.
\label{J7}
\end{equation}
The Lagrange multiplier $\lambda$ is obtained by convolution of (\ref{0A1}) with $U_j$; it has the following form:
\begin{equation}
\lambda =  U_j \left[\nabla_a {\cal J}^{aj}- I^j \right]  \,,  \label{0A309}
\end{equation}
where the quantities ${\cal J}^{aj}$ and $I^j$ are given by Eqs. (\ref{J1})- (\ref{J7}).

\subsection{Equations for the gravitational field}

The variation of the action (\ref{total2}) with respect to the metric
$g^{ik}$ yields the gravitational field equations, which can be presented in the following form
$$
R_{ik} - \frac{1}{2} R \ g_{ik}  -  \Lambda g_{ik} =
$$
\begin{equation}
=\lambda U_i U_k  + T^{({\rm U})}_{ik} + \kappa T^{({\rm h})}_{ik} +
\kappa T^{({\rm A})}_{ik} + \kappa T^{(\phi)}_{ik}
\,. \label{0Ein1}
\end{equation}
The term $T^{({\rm U})}_{ik}$ describes
the stress-energy tensor associated with the self-gravitation
of the vector field $U^i$:
$$
T^{({\rm U})}_{ik} =
\frac12 g_{ik} {\cal J}^{am}_{({\rm U})} \nabla_a U_m {+}
$$
$$
{+}\nabla^m \left[U_{(i}{\cal J}_{k)m}^{({\rm U})}\right] {-}
\nabla^m \left[{\cal J}_{m(i}^{({\rm U})}U_{k)} \right] {-}
\nabla_m \left[{\cal J}_{(ik)}^{({\rm U})} U^m\right]+
$$
$$
+C_1\left[(\nabla_mU_i)(\nabla^m U_k) {-}
(\nabla_i U_m \nabla_k U^m) \right] {+}
$$
\begin{equation}
{+}C_4 (U^a \nabla_a U_i)(U^b \nabla_b U_k) \,.
\label{5Ein1}
\end{equation}
As usual, the symbol $p_{(i} q_{k)}{\equiv}\frac12 (p_iq_k{+}p_kq_i)$
denotes the procedure of symmetrization.  The term $T^{({\rm h})}_{ik}$ can be presented similarly:
$$
T^{({\rm h})}_{ik} =
\frac12 g_{ik} {\cal J}^{am}_{({\rm h})} \nabla_a U_m {+}
$$
$$
{+}\nabla^m \left[U_{(i}{\cal J}_{k)m}^{({\rm h})}\right] {-}
\nabla^m \left[{\cal J}_{m(i}^{({\rm h})}U_{k)} \right] {-}
\nabla_m \left[{\cal J}_{(ik)}^{({\rm h})} U^m \right] +
$$
$$
+ \Psi^2_0 \phi^2 \left\{
h_1\left[(\nabla_m U_i)(\nabla^m U_k) {-}
(\nabla_i U_m)(\nabla_k U^m) \right] {+} \right.
$$
\begin{equation}
\left.+ h_4 (U^a \nabla_a U_i)(U^b \nabla_b U_k) \right\}\,.
\label{6Ein1}
\end{equation}
The quantity $T^{({\rm A})}_{ik}$ written as
$$
T^{({\rm A})}_{ik} = \Psi^2_0 \left\{\nabla_i \phi \nabla_k \phi {+} \frac12 g_{ik}\left[m^2_{({\rm A})} \phi^2 {-} \nabla_n \phi \nabla^n \phi \right]+\right.
$$
$$
\left.-
\frac12 \xi \left( \nabla_i \nabla_k {-} g_{ik}\nabla_m \nabla^m \right) \phi^2  \right. +
$$
\begin{equation}
\left.+\frac12 g_{ik}\left[h_{10} (U^m \nabla_m \phi)^2 {-} \alpha U^m \nabla_m (\phi^2) \right]\right\}
\label{7Ein1}
\end{equation}
describes the extended stress-energy tensor of the pseudoscalar field; the first line in this representation relates to the canonical part of the stress-energy tensor, the second line corresponds to the nonminimal coupling of the pseudoscalar field to the spacetime curvature, and the third line describes the contribution of the coupling of the axion field to the unit dynamic vector field, which contains the four-vector $U^k$ and does not include its covariant derivative.

The quantity
$$
T^{(\phi)}_{ik} =
\frac12 g_{ik} {\cal J}^{am}_{(\phi)} \nabla_a U_m {+}
$$
$$
+ \nabla^m \left[U_{(i}{\cal J}_{k)m}^{(\phi)}\right] {-}
\nabla^m \left[{\cal J}_{m(i}^{(\phi)} U_{k)} \right] {-}
\nabla_m \left[{\cal J}_{(ik)}^{(\phi)} U^m \right] +
$$
$$
+ \frac12 h_5 \Psi_0 \phi \epsilon^{abmn} (\nabla_b U_n)\left[4 \nabla_a U_{(i} g_{k)m} {-}g_{ik}(\nabla_aU_m) \right] +
$$
$$
{+} h_6 \Psi_0 \phi \epsilon^{abmn} \left\{ (\nabla_b U_n) U^s \nabla_s U_m \left[ g_{a(i} U_{k)} {-} \frac12 U_a g_{ik} \right] {+} \right.
$$
$$
\left.
+  U_a (\nabla_bU_n) U^s \nabla_s U_{(i} g_{k)m} +  \right.
$$
$$
\left.
+  U_a (U^s \nabla_s U_m) \nabla_b U_{(i} g_{k)n}  \right\}-
$$
$$
- \frac12 h_7 \Psi^2_0 g_{ik} U^m \nabla_m (U^n \nabla_n \phi^2) +
$$
$$
+ \frac12 h_8 \Psi^2_0 \left\{g_{ik}(U^a\nabla_a U^m) (\nabla_m \phi^2) - \right.
$$
$$
\left. -(\nabla_m U^m) \left[U_i (\nabla_k \phi^2)+ U_k (\nabla_i\phi^2)\right]
- \right.
$$
$$
\left.-\left[(U^m \nabla_m U_i)(\nabla_k\phi^2) + (U^m \nabla_m U_k)(\nabla_i \phi^2) \right] - \right.
$$
$$
\left.- \left[U_i U^m \nabla_m (\nabla_k \phi^2)+ U_k U^m \nabla_m(\nabla_i \phi^2) \right]
+ \right.
$$
$$
\left.
+(\nabla^m \phi^2) \left[U_i (\nabla_m U_k)+ U_k (\nabla_m U_i) \right] + \right.
$$
\begin{equation}
\left. + U_iU_k (\nabla^m \nabla_m \phi^2)\right\} \,,
\label{8Ein1}
\end{equation}
contains all possible sources described by cross - terms with the covariant derivative of the velocity four-vector, and the pseudoscalar field.

Compatibility conditions for the set of equations (\ref{0Ein1})
\begin{equation}
\nabla^k\left[ \lambda U_i U_k  + T^{({\rm U})}_{ik} + \kappa T^{({\rm h})}_{ik} +
\kappa T^{({\rm A})}_{ik} + \kappa T^{(\phi)}_{ik}
\right] = 0
\,, \label{compa1}
\end{equation}
are satisfied automatically on the solutions to master equations (\ref{ax1})-(\ref{ax4}), and (\ref{0A1})-(\ref{0A309}).

\subsection{Two remarks}

\subsubsection{Remark about the Khronon-like field in the Einstein-aether-axion theory}

The axionic extension of the Einstein-aether theory inherits the possibility to introduce the Khronon field $\Phi$ \cite{Khronon1}, when the unit velocity four-vector $U^i$ satisfies  the condition
\begin{equation}
U_i = \frac{1}{X} \nabla_i \Phi \,, \quad X^2 = \nabla_m \Phi \nabla^m \Phi 
\,. \label{Khr1}
\end{equation}
Let us remind that integrability conditions for (\ref{Khr1}) require that 
\begin{equation}
\nabla_{j} \nabla_{i} \Phi \equiv \nabla_{i} \nabla_{j} \Phi \ \ \Rightarrow  \ \ \nabla_{j} [X U_{i}] = \nabla_{i} [X U_{j}]
\,. \label{Khr2}
\end{equation} 
Convolution of (\ref{Khr2}) with $U^j$ yields
\begin{equation}
\nabla_{i} X = X DU_i + U_i DX \,,
 \label{Khr3}
\end{equation}
thus providing the first compatibility condition (\ref{Khr2}) for $U^i$ itself to be of the form 
\begin{equation}
\nablab_{[j} U_{i]} = 0 \ \ \Rightarrow \ \ \omega_{ji} =0 \,.
\label{Khr4}
\end{equation}
In other words, for the models, in which the vorticity tensor vanishes, i.e., $\omega_{ji} {=} 0$, the Khronon field exists. 
Extending this idea for the case of Einstein-aether-axion theory, one can assume that this Khronon field $\Phi$ could be linked with some function $F(\phi^2)$ depending on the square of the axion field $\phi$. In this sense the global time-like surface can be associated with the axionic dark matter distribution; such situation is admissible, for instance, for models describing Friedmann-type cosmology, static models with spherical symmetry, models with pp-wave symmetry, since  $\omega_{ji} {=} 0$ in all these cases. In nearest future, we hope to consider the problem of universal horizons in a static spherically symmetric Einstein-aether-axion model along the line proposed in the works \cite{Khronon1,Khronon2}.

\subsubsection{Remark about a Vortex-like field}

For models, in which $\omega_{ji} \neq 0$ and thus the Khronon field is not admissible, there exists another interesting possibility: one can introduce the pseudoscalar field $\psi$ by the relation
\begin{equation}
\omega_i  \equiv \epsilon_{ikpq} U^k \omega^{pq}= \Omega \nabla_i \psi  \,,
\label{Vor1}
\end{equation}
with some phenomenological constant $\Omega$ and the aether rotation (pseudo)four-vector $\omega_i$.
This pseudoscalar field can be indicated, e.g., as a Vortex-like field, and the most interesting case appears when the Vortex-like field $\psi$ is proportional to the axion field $\phi$.
Clearly, the compatibility conditions for (\ref{Vor1}) require that the field $\psi$ to be stationary, i.e., $D\psi{=}0$, and that the relation
\begin{equation}
2\omega^{*}_{k[i} \nabla_{j]}U^k = \eta_{pq[i} \nabla_{j]} \omega^{pq} 
\label{Vor2}
\end{equation}
is satisfied for the unit vector field itself.

\section{Application: The spatially isotropic and homogeneous cosmological model with dynamic aether and axionic dark matter}

\subsection{Reduced master equations}

In this Subsection we consider the master equations for the pseudoscalar and gravitational field for the symmetry associated with spatially isotropic,
 homogeneous cosmological model of the Friedmann type. We assume the metric to be of the form
\begin{equation}
ds^2 = dt^2 - a^2(t)[dx^2+dy^2+dz^2] \,,
\label{App1}
\end{equation}
with the scale factor $a(t)$ and the Hubble function $H(t){=}\frac{\dot{a}}{a}$. The dot denotes the derivative with respect to cosmological time $t$; we use the units with $c=1$.
We use the ansatz that the pseudoscalar and unit dynamic vector fields, $\phi$ and $U^i$, inherit the chosen symmetry. Mathematically, this requirement means that the Lie derivatives of the pseudoscalar and vector fields are equal to zero, when one calculates them along the Killing vectors $\left\{\xi^i_{(\alpha)} \right\}$, attributed to the chosen spacetime symmetry:
\begin{equation}
\pounds_{\xi^i_{(\alpha)}} \phi \equiv \xi^i_{(\alpha)} \partial_i \phi = 0 \,, \quad \alpha = 1,2,...6 \,,
\label{Kill1}
\end{equation}
\begin{equation}
\pounds_{\xi^i_{(\alpha)}} U^l \equiv \xi^i_{(\alpha)} \partial_i U^l - U^i \partial_i \xi^l_{(\alpha)}=0 \,.
\label{Kill2}
\end{equation}
It is well-known, that the symmetry associated with the Killing vectors  $\xi^i_{(1)} {=} \delta^i_1$, $\xi^i_{(2)} {=} \delta^i_2$, $\xi^i_{(3)} {=} \delta^i_3$, from the Abelian subgroup of the total 6-parameter symmetry group, provides the pseudoscalar and vector fields to be the functions of the cosmological time only, $\phi(t)$ and $U^i(t)$ (see (\ref{Kill1}) and (\ref{Kill2})). Using the Killing vectors describing the spatial rotations, we obtain from (\ref{Kill2})) that $U^1{=}U^2{=}U^3{=}0$. In other words, the velocity four-vector has to be of the form $U^i = \delta^i_0$, thus providing the absence of preferred spatial directions in the isotropic spacetime.

The covariant derivative $\nabla_i U_k$ in this case is characterized by vanishing acceleration four-vector, shear and vorticity tensors
\begin{equation}
DU^i = 0 \,, \quad \sigma_{mn}=0\,, \quad \omega_{mn}=0 \,.
\label{App2}
\end{equation}
Only the expansion scalar is nonvanishing, and one can write
\begin{equation}
\Theta = 3H(t) \,, \quad \nabla_i U_k =  \Delta_{ik} \ H(t) \,.
\label{App3}
\end{equation}
Our first task is to prove that for our ansatz the evolutionary equations for the unit vector field  (\ref{0A1})-(\ref{J7}) are satisfied identically; then we will consider the reduced equations for pseudoscalar and gravitational fields, and obtain exact solutions to these equations.

\subsubsection{Reduced equations for the unit vector field are satisfied identically}

Using the discussed above ansatz about the velocity four-vector, $U^i=\delta^i_0$, we can calculate explicitly all the necessary quantities:
\begin{equation}
{\cal J}^{aj}_{({\rm U})} = H(t) \left[\Delta^{aj}\left(C_1+3C_2+C_3 \right) + 3C_2U^a U^j \right]  \,,
\label{App3R}
\end{equation}
\begin{equation}
{\cal J}^{aj}_{({\rm h})} =  \Psi_0^2 \phi^2 H(t) \left[\Delta^{aj}\left(h_1{+}3h_2{+}h_3 \right) {+} 3h_2U^a U^j \right] \,,
\label{App4}
\end{equation}
\begin{equation}
{\cal J}^{aj}_{(\phi)} = 0\,, \quad I^j_{({\rm U})} =  0 \,,
\label{App5}
\end{equation}
$$
I^j_{(\phi)} =  \Psi^2_0 U^j \left\{D \phi \left(h_{10} D \phi {-}\alpha \phi \right) + \right.
$$
\begin{equation}
\left. +\frac12\kappa  \left[(h_7{+}h_8) DD\phi^2 + 3 H(t)(h_7{-}h_8) D \phi^2 \right] \right\} \,.
\label{App7}
\end{equation}
The four-vector $I^j$ is parallel to $U^j$. One can see that the divergence $\nabla_a {\cal J}^{aj}$ is also parallel to the velocity four-vector
\begin{equation}
\nabla_a {\cal J}^{aj} = 3U^j \left[{\cal B}(\dot{H}+3H^2)+ HD{\cal B} - H^2 {\cal A} \right] \,,
\label{App8}
\end{equation}
where the following notations are used:
$$
{\cal A} \equiv \left(C_1+3C_2+C_3 \right) + \kappa \Psi_0^2 \phi^2 \left(h_1+3h_2+h_3 \right)\,,
$$
\begin{equation}
{\cal B} \equiv C_2 + \kappa \Psi_0^2 \phi^2 h_2 \,.
\label{App9}
\end{equation}
These results are not surprising; since the spacetime of the Friedmann type is spatially isotropic  and homogeneous, there are no preferred space-like directions, and the four-vectors
$I^j_{(\phi)}$ and $\nabla_a {\cal J}^{aj}$ should be either parallel to $U^j$, or vanishing.
Thus, three of four basic evolutionary equations (\ref{0A1}) for the unit vector field are satisfied identically, and the last one defines the Lagrange multiplier (see (\ref{0A309})):
$$
\lambda(t) =
-3H^2 \left(C_1+3C_2+C_3 \right) + 3C_2 \left(\dot{H}+ 3H^2 \right) +
$$
$$
+ \kappa \Psi^2_0 \phi^2 \left[-3H^2 \left(h_1+3h_2+h_3 \right) + 3h_2 \left(\dot{H}+ 3H^2 \right) \right] -
$$
$$
- \kappa \Psi^2_0 \left[h_{10}{\dot{\phi}}^2 - \alpha \phi \dot{\phi} + (h_7+h_8)(\phi \ddot{\phi}+{\dot{\phi}}^2)
+ \right.
$$
\begin{equation}
\left.
+3H \phi \dot{\phi} (h_7-h_8-2h_2) \right] \,. \label{App11}
\end{equation}

\subsubsection{Reduced equation for the pseudoscalar (axion) field}

Since the vorticity tensor vanishes for the cosmological vector field $U^i=\delta^i_0$, we obtain immediately from (\ref{ax4}) that  ${\cal J}=0$, and that the evolutionary equation for the axion field takes the form
\begin{equation}
(1{-}h_{10})\left[\ddot{\phi} + 3H \dot{\phi} \right] + \phi {\cal H} = 0 \,.
\label{App12}
\end{equation}
The function ${\cal H}(t)$ is represented as follows:
$$
{\cal H}(t) \equiv  m^2_{({\rm A})} + 3\alpha H +  3 H^2 \left(h_1{+}3 h_2{+}h_3 {-}3h_7 \right)  -
$$
\begin{equation}
-3h_7 \dot{H} + 6 \xi \left(\dot{H}+2H^2 \right)    \,.
\label{App13}
\end{equation}
The function $\frac{{\cal H}(t)}{(1{-}h_{10})}$ plays the role of a squared effective mass of the axion coupled to the dynamic unit vector field and nonminimally coupled to the spacetime curvature.

\subsubsection{Reduced equations for the gravitational field}

The right-hand side of the evolutionary equations for the gravitational filed contains now the following four sources.
First, the stress-energy tensor describing the contribution of the unit dynamic vector field is
$$
T^{({\rm U})}_{ik} = - (C_1+3C_2+C_3) \Delta_{ik} \left(\dot {H} + \frac32 H^2 \right) +
$$
\begin{equation}
+ U_iU_k \left[\frac32 H^2(C_1+C_3)- 3C_2\left(\dot {H} + \frac32 H^2 \right) \right] \,.
\label{App14}
\end{equation}
Similarly, one can write the stress-energy tensor describing the interaction of the axion and vector fields:
$$
T^{({\rm h})}_{ik} = \Psi^2_0 \phi^2 \left\{ - (h_1+h_2+h_3) \Delta_{ik} \left(\dot {H} + \frac32 H^2 \right) + \right.
$$
\begin{equation}
\left. + U_iU_k \left[\frac32 H^2(h_1+h_3)- 3h_2\left(\dot {H} + \frac32 H^2 \right) \right]\right\} \,.
\label{App15}
\end{equation}
The extended stress-energy tensor of the pseudoscalar field is presented as follows:
$$
T^{({\rm A})}_{ik} {=} \Psi^2_0 \left\{\frac12 g_{ik}\left[m^2 \phi^2 {-} {\dot{\phi}}^2 (1{-} h_{10}){-} 2\alpha \phi \dot{\phi} \right]{+} U_i U_k {\dot{\phi}}^2 {+}\right.
$$
\begin{equation}
\left. + \xi \left[\Delta_{ik} (\phi \ddot{\phi}+ {\dot{\phi}}^2 + 2 H \phi \dot{\phi})+ 3H U_iU_k \phi \dot{\phi})\right]  \right\} \,.
\label{App16}
\end{equation}
The contribution of the interaction between axion and vector fields is described by the term
$$
T^{(\phi)}_{ik} =
-  \Psi^2_0 \left\{
\Delta_{ik} h_7 \left(\phi \ddot{\phi}+{\dot{\phi}}^2 \right) + \right.
$$
\begin{equation}
\left.  + U_iU_k\left[(h_7+h_8)\left(\phi \ddot{\phi} + {\dot{\phi}}^2 \right) + 3Hh_8 \phi \dot{\phi} \right]
\right\} \,.
\label{App17}
\end{equation}

\subsubsection{A resume: The key equation for the gravity field}

As usual, only one of two non-trivial equations for the gravity field is independent for the symmetry associated with the Friedmann-type model; the second equation is the differential consequence, since the compatibility conditions are satisfied identically for the solutions to the axion field equation (\ref{App12}). This key equation can be written in the following form:
$$
\frac{1}{\kappa \Psi^2_0}\left\{3H^2 \left[1+ \frac12 \left(C_1{+}3C_2{+}C_3 \right) \right] {-} \Lambda \right\}  =
$$
$$
- \frac32 \phi^2  H^2 \left(h_1{+}3h_2{+}h_3 \right) {-} 2 (h_7{+}h_8)(\phi \ddot{\phi}{+}{\dot{\phi}}^2) -
$$
\begin{equation}
-3H \phi \dot{\phi} (h_7{-}2h_2{-}\xi)
{+}
\frac12 \left[m^2_{({\rm A})} \phi^2 {+} {\dot{\phi}}^2(1 {-} h_{10})  \right] \,.
\label{App18}
\end{equation}
In the master equation for the pseudoscalar field (\ref{App12}), (\ref{App13}) there are eight parameters regulating the behavior of solutions:
$h_{10}$, $m_{({\rm A})}$, $\alpha$, $h_1$, $h_2$, $h_3$, $h_7$, $\xi$. In addition, the key equation for the gravity field (\ref{App18}) contains
seven parameters: $\Lambda$, $\kappa$, $\Psi_0$, $C_1$, $C_2$, $C_3$, $h_8$. Of course, these fifteen parameters have different status. The parameter $\kappa$ is constructed from
the well-known fundamental constants and can not be changed. The cosmological constant $\Lambda$ is the main guiding parameter; for the $\Lambda$CDM theory the cosmological constant is responsible for the Dark Energy (see, e.g., \cite{DE1,DE2,DE3}). Two parameters, the axion mass $m_{({\rm A})}$ and the constant of the axion-photon coupling $\Psi_0$, are the constants predetermining the properties of the axionic Dark Matter; these parameters are not fixed, but there are constraints for them (see, e.g., \cite{ADM5,ADM8}). Three Jacobson's constants from the four ones introduced in the Einstein-aether theory, namely, $C_1$, $C_2$, $C_3$, also are not fixed, but constrained; in our cosmological model these constants form only one guiding parameter $C {=}C_1{+}3C_2{+}C_3$. The parameter $\xi$ is the pseudo-analog of the well-known constant of nonminimal coupling of the scalar field; now there are no fixed estimations for it. Parameters $h_{10}$, $\alpha$, $h_1$, $h_2$, $h_3$, $h_7$, $h_8$ are novel. Similarly to the Jacobson's parameters, the constants $h_1$, $h_2$, $h_3$ form only one guiding parameter $h {=}h_1{+}3h_2{+}h_3$. We assume that the parameters $h_{10}$, $\alpha$, $h$, $h_7$, $h_8$ are connected by  some relationships; we will discuss them below.

Now we are ready to consider examples of exact solutions to the equations (\ref{App12}) and (\ref{App18}).

\subsection{Example 1: Exact solution of the de Sitter type}

Representing this first example, we pose a question: whether the solution exists, for which the Hubble function $H(t)$ is constant, say, $H(t){=}H_0$. The answer is yes. Indeed,
let the coupling parameters of the Einstein-aether-axion model be chosen according to the following scheme:
$$
h_{10} = 1-2\nu \,,
\quad
h_1+h_3 = 3\nu - \frac12 \xi - \frac{\alpha}{2H_0} \,,
$$
$$
h_2 = \frac{m^2_{({\rm A})}}{9H^2_0} + \frac{\alpha}{6H_0}-\frac12 \nu + \frac16 \xi \,,
$$
$$
h_7 = \frac{2 m^2_{({\rm A})}}{9H^2_0} + \frac{\alpha}{3H_0} +\frac43 \xi\,,
$$
\begin{equation}
h_8 = \frac12 \nu -\frac{2 m^2_{({\rm A})}}{9H^2_0} - \frac{\alpha}{3H_0} -\frac43 \xi\,.
\label{App21}
\end{equation}
Then the master equations of the model are reduced to
\begin{equation}
\ddot{\phi} + 3H_0 \dot{\phi} + \frac94 H^2_0 \phi  = 0 \,,
\label{App22}
\end{equation}
\begin{equation}
3H^2_0 \left[1+ \frac12 \left(C_1+3C_2+C_3 \right) \right] = \Lambda  \,.
\label{App23}
\end{equation}
The solution to the Hubble function is indeed constant
\begin{equation}
H_0 = \sqrt{\frac{\Lambda}{3}} \left[1+ \frac12 \left(C_1+3C_2+C_3 \right) \right]^{-\frac12} \,,
\label{App24}
\end{equation}
the scale factor being of the de Sitter form $a(t){=}a_0 e^{H_0 t}$.
The function $\phi(t)$, the corresponding solution for the axion field
\begin{equation}
\phi(t) = t e^{-\frac32 H_0 t} \dot{\phi}(0) \,,
\label{App25}
\end{equation}
is chosen to satisfy the initial conditions $\phi(0){=}0$, $\dot{\phi}(0)\neq 0$. The function $\phi(t)$ grows with time, reaches the maximal value $\phi_{({\rm max})} {=} \frac{2}{3eH_0} \dot{\phi}(0)$ at $t_{*}{=}\frac{2}{3H_0}$, and then tends asymptotically to zero at $t \to \infty$.

\subsection{Example 2: Exact solutions with asymptotically de Sitter behavior}

Now we consider the following relationships between the coupling parameters:
$$
\left(h_1{+}3 h_2{+}h_3 \right) =  2\xi - 3 (h_7-2\xi)^2 \,,
$$
\begin{equation}
h_{10} = 0 \,, \quad \alpha =0 \,, \quad m^2_{({\rm A})} = 0 \,.
\label{App26}
\end{equation}
For this case the equation describing the axion field
\begin{equation}
\ddot{\phi} + 3H \dot{\phi} = -\sigma \phi \left[\dot{H}+ H^2\left(3-\sigma \right) \right] \,,
\label{App27}
\end{equation}
contains one effective guiding parameter
\begin{equation}
\sigma \equiv {-} 3(h_7{-}2\xi) \,.
\label{sigma}
\end{equation}
One can check directly, that for arbitrary $\dot{H}$ and $H^2$ there exists the following exact solution of (\ref{App27}):
\begin{equation}
\phi(t) = \phi(t_0) \left[\frac{a(t)}{a(t_0)} \right]^{-\sigma}  \,.
\label{App28}
\end{equation}
We assume that the parameter $\sigma$ is positive, $\sigma>0$, providing the axion field (\ref{App28}) to tend asymptotically to zero at $a\to \infty$.
For this solution
\begin{equation}
\dot{\phi} =- \sigma H(t) \phi(t) \,, \quad  \ddot{\phi} = -\sigma \phi(t) \left[\dot{H}- \sigma H^2 \right]\,,
\label{App29}
\end{equation}
and thus the key equation for the gravity field (\ref{App18}) can be written in the form
$$
\sigma (h_7{+}h_8) x \frac{d}{dx} H^2(x)+  \frac{\Lambda  x^{\sigma}}{\kappa \Psi^2_0 \phi^2(t_0)}
=
$$
$$
= 3H^2 \left\{\left[\xi(1{-}2\sigma) {+} \frac43 (h_7{+}h_8) \sigma^2\right] +  \right.
$$
\begin{equation}
\left. + \frac{x^{\sigma}}{\kappa \Psi^2_0 \phi^2(t_0)} \left[1{+} \frac12 \left(C_1{+}3C_2{+}C_3 \right) \right] \right\} \,.
\label{App30}
\end{equation}
For the sake of convenience we introduced here a new dimensionless variable $x {=} \frac{a(t)}{a(t_0)}$, for which
\begin{equation}
\frac{d}{dt} = x H \frac{d}{dx} \,, \quad \phi(x) = \phi(t_0) x^{-\sigma} \,.
\label{App31}
\end{equation}
The initial time moment $t{=}t_0$ relates to $x{=}1$; for an expanding Universe $x \geq 1$. Clearly, the sum $h_7{+}h_8$, which appears in front of the derivative $\frac{d}{dx} H^2(x)$, introduces a new effective guiding parameter of the model; below we distinguish the submodels with
$h_7{+}h_8{=}0$ and $h_7{+}h_8 \neq 0$.

\subsubsection{The model with $h_7{+}h_8=0$}

When $h_7{+}h_8{=}0$, the solution to (\ref{App30}) has the form
\begin{equation}
H(x)= \frac{H_0}{\sqrt{1+ \Gamma x^{-\sigma}}} \,,
\label{App32}
\end{equation}
where $H_0$ is given by (\ref{App24}), and the parameter $\Gamma$ is
\begin{equation}
\Gamma \equiv \frac{\xi(1{-}2\sigma)\kappa \Psi^2_0 \phi^2(t_0)}{\left[1{+} \frac12 \left(C_1{+}3C_2{+}C_3 \right) \right]} \,.
\label{App33}
\end{equation}
We assume that $\Gamma > {-}1$, providing the initial value  $H(1)= \frac{H_0}{\sqrt{1+ \Gamma}}$ to be the real quantity; with this condition $H(x)$ is real for $t\geq t_0$ ($x\geq 1$).
In order to reconstruct the function $a(t)$ we have to use the integral
\begin{equation}
t-t_0 = \int_1^{\frac{a(t)}{a(t_0)}} \frac{dx}{x H(x)} \,,
\label{App34}
\end{equation}
which can be now written as
\begin{equation}
H_0 (t-t_0) = \int_1^{\frac{a(t)}{a(t_0)}} \frac{dx}{x} \sqrt{1+ \Gamma x^{-\sigma}} \,.
\label{App35}
\end{equation}
Integration in (\ref{App35}) yields
$$
\sigma H_0 (t-t_0) = 2 \left[\sqrt{1+ \Gamma}- \sqrt{1+ \Gamma x^{-\sigma}} \right] +
$$
\begin{equation}
+\log{\left|\frac{(\sqrt{1+ \Gamma}-1)(\sqrt{1+ \Gamma x^{-\sigma}}+1)}{(\sqrt{1+ \Gamma}+1)(\sqrt{1+ \Gamma x^{-\sigma}}-1)} \right|} \,.
\label{App36}
\end{equation}
When $\xi (1{-}2\sigma) \to 0$, i.e., $\Gamma \to 0$, we obtain from (\ref{App36}) that the right-hand side tends to $\sigma \log{x}$, and thus we recover the de Sitter solution $a(t)=a(t_0)e^{H_0(t-t_0)}$. Clearly, the same scale factor can be obtained, when $x \to \infty$, i.e., this solution demonstrates the de Sitter asymptote for arbitrary $\Gamma>-1$.

\subsubsection{The model with $h_7{+}h_8\neq 0$, $\Lambda=0$}

For the model of this type we can rewrite the key equation (\ref{App18}) as follows:
\begin{equation}
x \frac{d}{dx} \log{H} = \omega_1 + \sigma \omega_2 x^{\sigma} \,,
\label{App37}
\end{equation}
where
\begin{equation}
\omega_1 \equiv 2 \sigma + \frac{3\xi(1{-}2\sigma)}{2\sigma(h_7{+}h_8)} \,,
\label{App38}
\end{equation}
\begin{equation}
\omega_2 \equiv \frac{3\left[1{+} \frac12 \left(C_1{+}3C_2{+}C_3 \right) \right]}{2\sigma^2 (h_7{+}h_8)\kappa \Psi^2_0 \phi^2(t_0)} \,.
\label{App39}
\end{equation}
The solution for the Hubble function $H(x)$ is now
\begin{equation}
H(x) = H(t_0) \ x^{\omega_1} e^{\omega_2\left(x^{\sigma}-1 \right)} \,.
\label{App40}
\end{equation}
The scale factor can be found using the integral
\begin{equation}
H(t_0) (t-t_0) e^{-\omega_2} = \int_1^{\frac{a(t)}{a(t_0)}} \ dx  x^{-(\omega_1{+}1)} e^{-\omega_2 x^{\sigma}} \,.
\label{App409}
\end{equation}
The behavior of the functions $H(x)$ and $a(t)$ depends on the sign of the guiding parameter  $\omega_2$ (it coincides with the sign of the parameter $h_7{+}h_8 \neq 0$).
Let us consider two principal cases.

\vspace{3mm}
\noindent
{\it (i)} $\omega_2>0$.

\noindent
When $a(t) \to \infty$, the integral in the right-hand side of (\ref{App409}) converges and gives a finite quantity. This means that during a finite interval of time $t_0<t<t_{*}$ the scale factor reaches infinite value; clearly, we deal with the case indicated as Big Rip (see, e.g., \cite{O1,O2}). The Hubble function also reaches infinity at $t{=}t_{*}$. The guiding parameter $\omega_1$ is responsible for extrema of the function $H(x)$.
Since $\omega_2>0$, the extremum of the function $H(x)$ given by (\ref{App40}) exists, if  $\omega_1<0$; in this case we deal with a minimum  at
\begin{equation}
x=x_{*} \equiv \left[\frac{|\omega_1|}{\sigma |\omega_2|} \right]^{\frac{1}{\sigma}} \,,
\label{App41}
\end{equation}
with the minimal value
\begin{equation}
H_{({\rm min})}= H(x_{*}) = H(t_0) \left[\frac{|\omega_1|}{\sigma |\omega_2|} \right]^{-\frac{|\omega_1|}{\sigma}} e^{\frac{|\omega_1|}{\sigma}-|\omega_2|} \,.
\label{App421}
\end{equation}

\vspace{3mm}
\noindent
{\it (ii)} $\omega_2<0$.

\noindent
For negative parameter $\omega_2$ the integral in (\ref{App409}) diverges at $a(t) \to \infty$. This means that infinite value of the scale factor can be reached only during the infinite time interval. According to (\ref{App40}) $H(x\to \infty) \to 0$, the function $H(x)$ being monotonic, when $\omega_1<0$.  When $\omega_2<0$ and $\omega_1>0$, there is a maximum at
$x=x_{*}$, and
\begin{equation}
H_{({\rm max})}= H(x_{*}) = H(t_0) \left[\frac{|\omega_1|}{\sigma |\omega_2|} \right]^{\frac{|\omega_1|}{\sigma}} e^{-\frac{|\omega_1|}{\sigma}+|\omega_2|} \,.
\label{App42}
\end{equation}
For an illustration, one can consider the explicit case with $\omega_1{=} {-}\sigma<0$, and $\omega_2<0$, which is possible, when the sum $h_7+h_8$ is negative and
\begin{equation}
\left|h_7+h_8 \right|= \frac{\xi(1{-}2\sigma)}{2\sigma^2} >0  \,.
\label{App44}
\end{equation}
For this special combination of coupling parameters the scale function is described as follows:
\begin{equation}
\frac{a(t)}{a(t_0)}= \left\{1{+} \frac{1}{|\omega_2|}\log{\left[1{+} \sigma |\omega_2|H(t_0)(t{-}t_0) \right]} \right\}^{\frac{1}{\sigma}} \,.
\label{App45}
\end{equation}
The Hubble function (\ref{App40}) vanishes asymptotically, but the scale factor tends to infinity. The behavior (\ref{App45}) can be indicated as  a Pseudo Rip.

Also, one can mention some special case, when $\omega_2{=}0$, which is possible for $\Lambda{=}0$ and $\left(C_1{+}3C_2{+}C_3 \right)={-}2$. The Hubble function $H(x)$ is now
the power-law one, $H(x) {=} H(t_0) \ x^{\omega_1}$, and the scale factor is presented by
\begin{equation}
a(t) = a(t_0) \left[1{-} \omega_1 H(t_0)(t{-}t_0) \right]^{-\frac{1}{\omega_1}} \,.
\label{App456}
\end{equation}
Clearly, when $\omega_1>0$, we deal with a Big Rip; the scale factor and the Hubble function reach infinity at the moment $t=t_{*}=t_0{+}\frac{1}{|\omega_1|H(t_0)}$.
When $\omega_1<0$, we obtain a Pseudo Rip, since at $t \to \infty$ $a \to \infty$ as $a \propto t^{\frac{1}{|\omega_1|}}$, but $H(t)$ tends to zero.

\subsubsection{The model with $h_7{+}h_8\neq 0$, $\Lambda\neq 0$}

For this general case the solution for the scale factor can not be presented in explicit form, and we consider one illustration only. Let the parameters be coupled as follows:
\begin{equation}
h_7+h_8 =-\frac{3\xi (1-2\sigma)}{4\sigma^2} \,.
\label{App46}
\end{equation}
Then the key equation for the Hubble functions takes the form
\begin{equation}
x^{1-\sigma} \frac{d}{dx} H^2 = - \sigma \omega \left(H^2-H_0^2 \right) \,,
\label{App47}
\end{equation}
where $H_0$ is given by (\ref{App24}) and the parameter $\omega$ is defined as
\begin{equation}
\omega \equiv \frac{4\left[1{+} \frac12 \left(C_1{+}3C_2{+}C_3 \right) \right]}{\xi (1{-}2\sigma) \kappa \Psi^2_0 \phi^2(t_0)} \,.
\label{App48}
\end{equation}
The integration of (\ref{App47}) gives the Hubble function in explicit form
\begin{equation}
H(x) = \sqrt{H^2_0 + \left[H^2(t_0){-}H_0^2 \right] \exp\left\{\omega\left(1{-}x^{\sigma}\right)\right\}} \,.
\label{App49}
\end{equation}
The scale factor can be found from the implicit equation
\begin{equation}
\sigma H_0 (t-t_0) = \int_1^{\left(\frac{a(t)}{a(t_0)}\right)^{\sigma}}\frac{dz}{z \sqrt{1+ {\cal F} \exp\left\{-\omega z\right\}}} \,,
\label{App59}
\end{equation}
where the auxiliary parameter ${\cal F}$ is defined as
\begin{equation}
{\cal F} =  \left[\frac{H^2(t_0)}{H^2_0}{-}1 \right] e^{\omega}\,.
\label{App597}
\end{equation}
One can distinguish three interesting cases.

\vspace{3mm}
\noindent
{\it (i)} $\omega>0$.
\noindent
The Hubble function (\ref{App49}) starts with the value $H(t_0)$ and reaches asymptotically the value $H_0$ at $x \to \infty$.  The integral (\ref{App59}) diverges logarithmically at $a\to \infty$, providing the asymptotic behavior to be of the de Sitter-type $a(t) \to a_0 e^{H_0 t}$.

\vspace{3mm}
\noindent
{\it (ii)} $\omega<0$, $H^2(t_0)>H^2_0$.

\noindent
In this case the integral in (\ref{App59}) converges at $a\to \infty$, and thus we deal with a Big Rip, since the infinity is reached during the finite interval of cosmological time.
The Hubble function tends asymptotically to infinity as
\begin{equation}
H(x) \ \to \  \sqrt{H^2(t_0){-}H_0^2}  \ \exp\left\{\frac12|\omega|x^{\sigma} \right\} \,.
\label{App50}
\end{equation}

\noindent
{\it (iii)}  $H(t_0)=H_0$.

\noindent
Clearly, this is a de Sitter-type solution, since $H(x){=}H_0$, and $ a(t) {=} a(t_0) \exp\{H_0 (t{-}t_0)\}$.

The case $\omega<0$, $H^2(t_0)<H^2_0$ is non-physical, since there is a point, in which the function $H^2(x)$ changes the sign, and the Hubble function becomes a pure imaginary quantity.

\section{Discussion}

The full-format phenomenological model of interactions between the gravitational field, unit dynamic vector field and pseudoscalar (axion) field is established as a model of the
second order in the nomenclature of the Effective Field theory. Master equations of this version of the Einstein-aether-axion theory are derived and presented by Eqs. (\ref{0Ein1})-(\ref{8Ein1}),
(\ref{21})-(\ref{0A309}), (\ref{ax1})-(\ref{ax4}). This phenomenological theory includes seventeen coupling parameters in addition to the standard constants $c$, $G$ and $\Lambda$ (see (\ref{total2}) for the total action functional). In the spatially isotropic and homogeneous cosmological model only nine parameters (appearing with the expansion scalar $\Theta {=}3H(t)$) from these seventeen constants are shown to be activated;  other eight remain hidden parameters because of the symmetry of this model.

The principle known as "Occam's razor" requires to reduce the wide set of phenomenological parameters to a minimal number of key coupling constants. What instruments are in hands of theorists to diminish the number of phenomenological parameters? One of the instruments is the removal of singularities in the field configurations. For instance, the requirement of regularity of the gravitational field at the center of nonminimal black holes reduces the number of coupling constants from three to one \cite{BLZ2016}. The electric field at the center of nonminimal charged monopole, star or black hole can take finite value $E(0) \neq \infty$, when the coupling parameters satisfy two specific conditions (see \cite{BZ2015}).

We can use the same idea in the Einstein-aether-axion theory established in this work, in order to reduce the number of coupling constants. For instance, when we considered the application of the theory to cosmology, we have found that the Big Rip scenario can be realized for some set of coupling parameters. The Big Rip can be classified as a specific (late-time) singularity. We would like to avoid the Big Rip scenario in our cosmological models, which  describe  interactions between axionic dark matter, $\Lambda$-type dark energy and unit dynamic vector field associated with aether. For this purpose one can do the following procedure. One can introduce, for instance, five relationships (\ref{App21}) excluding the parameters $h_{10}$, $h_1{+}h_3$, $h_2$, $h_7$, and $h_8$, to guarantee that the Universe expands according to the de Sitter scenario with the Hubble constant (\ref{App24}) containing the cosmological constant $\Lambda$ and the composite Jacobson's parameter $C_1{+}3C_2{+}C_3$.

For a static model with spherical symmetry, the decomposition of the covariant derivative of the aether velocity four-vector includes the non-vanishing radial component of the acceleration four-vector $DU^i$. This means that other coupling constants (say, $h_8$), appeared in front of $DU^i$ in the total action functional (\ref{total2}),  will be activated in the Einstein-aether-axion model with such symmetry. Similarly, the non-vanishing shear tensor appears, when we deal with the model possessing a plane-wave symmetry \cite{AB2016}. The vorticity tensor becomes an important player in the Einstein-aether-axion theory, if one deals with a rotating Universe or rotating stars. We hope to analyze  the models with the corresponding symmetries in the nearest future to avoid singularities of the gravitational, unit vector and pseudoscalar fields, and to constrain the set of phenomenologically introduced coupling parameters.

We also keep in mind various applications of the Einstein-Maxwell-aether-axion theory, which is the next natural step in the extension of the theory established in this paper. However, this extension requires to use invariant terms of the third and fourth order according to the nomenclature of the Effective Field theory, so, it will be a new interesting work.

\acknowledgments{The work was partially supported by
the Program of Competitive Growth of Kazan Federal University, and by the Russian
Foundation for Basic Research (Grants RFBR No.~15-52-05045 and No.~14-02-00598). }

\end{document}